\documentclass{article}
\usepackage{blindtext}
\usepackage{graphicx}
\usepackage{subfigure}
\usepackage{cite}
\usepackage{epstopdf}
\DeclareGraphicsExtensions{.eps}

\usepackage{spconf, pifont, enumerate, amsmath, setspace, mathtools, amssymb, amsthm, subfigure, cite, graphicx,todonotes, multicol}

\theoremstyle{definition}

\theoremstyle{remark}

\theoremstyle{plain}

\begin{document}

\title{E\MakeLowercase{lastic} P\MakeLowercase{ath}2P\MakeLowercase{ath}: Automated morphological classification of neurons by elastic path matching.}

\name{Tamal Batabyal,~Scott T. Acton,~\emph{Fellow, IEEE}}

\address{Virginia Image and Video Analysis Laboratory,\\
	Department of Electrical \& Computer Engineering, University of Virginia, USA\\
	Charlottesville, VA 22904 U.S.A.}

\maketitle

\begin{abstract}
In the study of neurons, morphology influences function. The complexity in the structure of neurons poses a challenge in the identification and analysis of similar and dissimilar neuronal cells. Existing methodologies carry out structural and geometrical simplifications, which substantially change the morphological statistics. Using digitally-reconstructed neurons, we extend the work of Path2Path as \emph{ElasticPath2Path}, which seamlessly integrates the graph-theoretic and differential-geometric frameworks. By decomposing a neuron into a set of paths, we derive graph metrics, which are path concurrence and path hierarchy. Next, we model each path as an elastic string to compute the geodesic distance between the paths of a pair of neurons. Later, we formulate the problem of finding the distance between two neurons as a path assignment problem with a cost function combining the graph metrics and the geodesic deformation of paths. ElasticPath2Path is shown to have superior performance over the state of the art.        
\end{abstract}


\section{Introduction}  
\label{Intro}
In 1899, Santiago Ram\'{o}n y Cajal~\cite{y1972histologie} postulated in his seminal work that the shape of a neuron affects functionality. Over the last three decades, numerous research studies have been published to categorize and rigorously analyze the structural and geometric features of neurons in order to explore the functional correspondence. However, the complexity in the morphology, such as highly ramified arborization patterns~\cite{gillette2015topological, gillette2015topological1, kanari2017topological}, size~\cite{brown2008quantifying}, symmetry, and fragmentation~\cite{wan2015blastneuron} makes the analysis overwhelmingly difficult. Here, we seek to extend previous morphological image analysis work~\cite{acton2001fast, acton2000area, bosworth2003morphological, sarkar2013shape} in a graph theoretic framework to provide an elastic measure of distance between two complex shapes, i.e., the shapes of neurons.

A bottleneck in achieving improved performance in categorization stems from the data analysis. Morphological analysis of neurons from images are adversely affected by the existing problems in image processing. Neuromorpho.org~\cite{ascoli2007neuromorpho} provides a suitably alternative data modality to analyze the shape of a neuron sampled at contiguous 3D locations with respect to a fixed coordinate system. Owing to the discrete nature of the data, a neuron can be modeled as a graph, which helps in the localized extraction of features at multiple scales and the performance of global comparison. 

As a state-of-the-art approach, Gillette \emph{et al.} decomposed a neuron as a sequence of branches as local entities, and compared a pair of neurons using alignment of such branches. BlastNeuron~\cite{wan2015blastneuron} offered a framework which extracts structurally invariant features and geometric features to compare two neurons using a dynamic programming approach. However, this approach is preprocessing-dependent, which might alter the morphological statistics of a neuron. In another work, the Petilla terminology~\cite{ascoli2008petilla} ontologizes neuron features for a large volume of neuronal data. Graph-based registration-free approaches, such as NeuroSoL~\cite{batabyal2017neurosol} attempt to integrate a subset of local features to perform a global comparison. However, the alignment of such features for a pair of neurons is of non-polynomial computational complexity. A recent work, NeuroBFD~\cite{batabyal2018neurobfd} introduced a size and registration independent approach, which extracts empirical conditional distributions of three features, which are bifurcation angle, branch fragmentation, and spatial density. 
\begin{figure}[t]
\vspace{-0.2cm}
\centering
	\subfigure[]{\includegraphics[width=4.2cm,height=2.6cm]{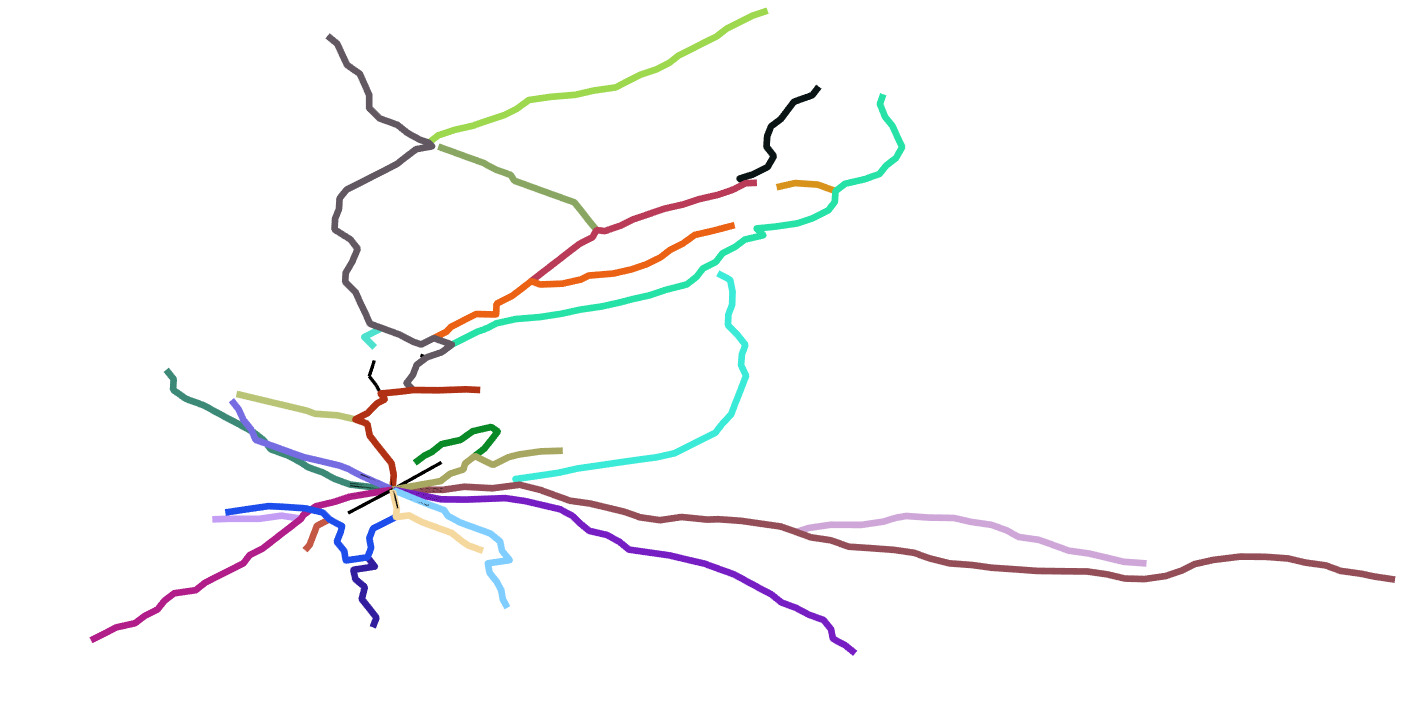}}
	\subfigure[]{\includegraphics[width=4.2cm,height=2.6cm]{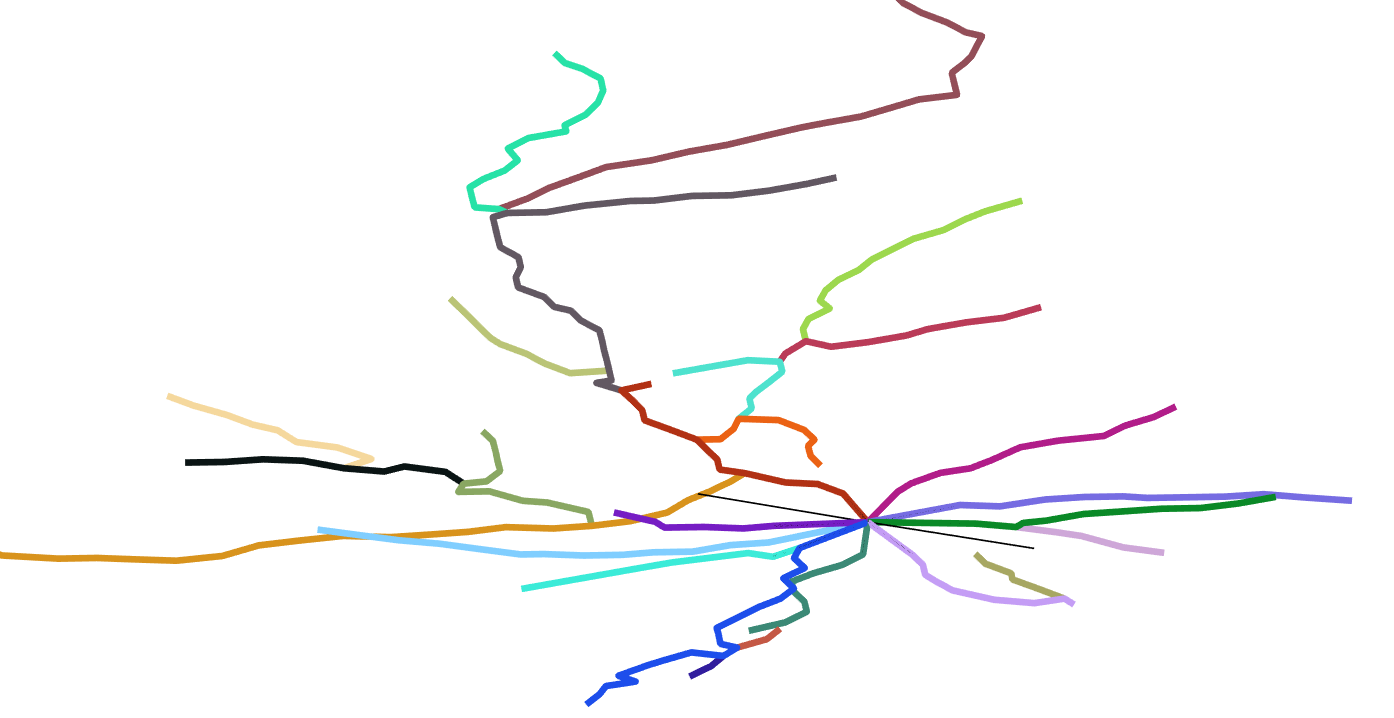}}	
	\caption{\small The path correspondences are shown in the corresponding colors between a pyramidal cell (463 3D locations, 28 paths) and another pyramidal cell (299 3D locations, 26 paths). The distance between the two neurons is 101.68 as computed by Elastic Path2Path.}
\label{figfig}
\vspace{-0.6cm}
\end{figure}
To represent the topological arbor, which is believed to be an indicator of complex neuronal computation~\cite{cuntz2007optimization}, the works of Path2Path~\cite{basu2011path2path} and topological morphology descriptor (TMD)~\cite{kanari2017topological} primarily considered the assembly of paths to compare two neurons. In Path2Path, the authors defined and extracted the path concurrence and path hierarchy values. On the other hand, TMD encodes birth and death instances of each path in a neuron tree into a persistence diagram. However, no appropriate distance measure is available for comparing persistence diagrams. 

Although authors in Path2Path~\cite{basu2011path2path} defined a distance measure using physical 3D coordinates, concurrence and hierarchy values to compare two paths, this approach has several technical drawbacks. Due to the greedy nature of the path assignment algorithm in order to compare two neurons, multiple paths of one neuron can be matched to a single path of the other neuron. We address this problem in section~\ref{p2pm}, and provide an optimal solution for bijective path matching as shown in Fig.~\ref{figfig}. Path2Path computes the physical distance between two paths without considering the nature of the manifold on which the two paths exist. In section~\ref{emorph}, we incorporate differential geometry by considering each path as an elastic open curve and provide a suitable transformation of the elastic curves. The geodesic is then computed to find the physical distance between two paths. Another problem associated with Path2Path is the resampling procedure, which changes the 3D locations on a path and adds other samples in between in an arbitrary order. This resampling creates a problem in the interpolation of concurrence and path hierarchy values. In addition, the resampling routine disrupts the `ordered' nature of each path emanating from the root. We briefly introduce our linear time iterative resampling scheme as given in section~\ref{emorph}.      
\begin{figure}[t]
\vspace{-0.2cm}
\centering
	\includegraphics[width=8.4cm,height=2.7cm]{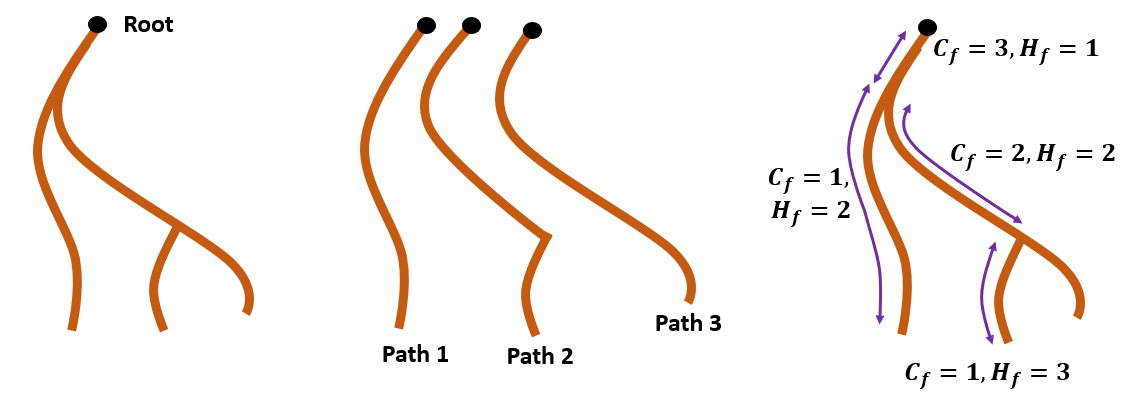}
	\caption{\small (a) This figure shows a schematic of a neuron as a rooted tree with three paths from the root to the leaf vertices. The path concurrence $C_f$ and hierarchy values, $H_f$ are depicted. The 3D locations that belong to a path segment are assigned the $C_f$ and $H_f$ values of that path segment.}
\label{fig1}
\vspace{-0.6cm}
\end{figure}

\textbf{Why ElasticP2P?}\hspace*{.2cm} Path2Path offered a computational framework for the quantitative assessment of the categorization of neuron cell types. Path concurrence and hierarchy are the two customized features used in~\cite{basu2011path2path} to describe a path. This framework has the flexibility to incorporate other customized features. The addition of $elasticity$ to Path2Path has an enormous impact, because it helps provide a framework to morph one neuron into another. This facility is entirely missing in existing approaches. \emph{The visualization aids in the validation of the correspondence of paths between two neurons for meaningful comparison}. The path morphing visually verifies the reasonable choice of the customized features and distance measure in elastic Path2Path. In contrast, conventional methods rely on performance scores to evaluate the strength of the features.


\section{Graph theory}
A graph is defined as a triplet $\mathcal{G} = \{\mathcal{V},\mathcal{E},\mathcal{W}\}$, where $\mathcal{V}$, $\mathcal{E}$, and $\mathcal{W}$ are the set of vertices (or nodes), edges (or links), and edge-weights respectively. A graph is said to be \emph{simple} if the graph does not contain multiple edges between any pair of vertices and any self-loop on a vertex. An \emph{undirected} graph has no direction of its edges. A path of length $k$ of $\mathcal{G}$ is defined as a sequence of $k$ distinct vertices. If the start and end vertices of a path are same, the path is termed as a \emph{loop}. A \emph{tree} is a graph with no loop. A \emph{connected} graph is the one in which there exists at least one path between any pair of vertices. The path between an arbitrary pair of vertices of a tree is unique.   

\section{methodology}
\subsection{Neuron as a graph}
We model a neuron with a simple, connected, undirected tree $\mathcal{G}$ with a designated root node. A neuron $\mathcal{G}$ with $n$ dendritic terminals can be decomposed into $n$ paths $f_i;~\{1,2,...,n\}$. $f_i$ is a continuous function for each path such that $f_i:[0,1]\longrightarrow \mathcal{R}^3$ with $f(0) = (0,0,0)$. Let $\Gamma$ be the set of all such $f_i$s, which is a linear subspace of the classical Wiener space. For numerical computation, $f_i$ is finitely sampled and each sample is treated as a vertex of the neuron graph. 

The concept of \emph{path concurrence}, $C_{f_i}$ of a vertex originates from counting the number of times a given vertex is revisited in all $f_i$s. If $C_{f_i}(t_s)=k;~t_s\in[0,1],~k\in\mathcal{N}$, the vertex at $t_s$ on the path $f_i$ is shared among $k$ paths of the neuron, $\mathcal{G}$. As given in~\cite{basu2011path2path} the path concurrence value can be mathematically represented by
\begin{eqnarray}
C_{f_i}(t) = k; ~t\in[0,1],~j_1,j_2,\dots,j_k\in\{1,2,...,n\}\nonumber\\
f_{i}(t) = f_{j_1}([0,1])\cap f_{j_2}([0,1])\cap \dots\cap f_{j_k}([0,1]).
\end{eqnarray}
After the computation of the path concurrence values at all the vertices, authors in~\cite{basu2011path2path} introduced the concept of path hierarchy values. The path hierarchy value of a vertex on a path $f_{i}$, $H_{f_i}(t_s)$ is the number of times the concurrent paths do not visit the vertex while traversing from the root node to the dendritic terminal of $f_i$.

In~\cite{basu2011path2path}, each vertex on an arbitrary path, $f_i$ therefore has 3D location, path concurrence value $C_{f_i}(t_s)$ and path hierarchy value $H_{f_i}(t_s)$ as shown in Fig.\ref{fig1}. The distance metric between any two paths $f_i$ and $f_j$ was given in  \cite{basu2011path2path} as 
\begin{eqnarray}
\label{p2pmetric}
D\big(f_i,f_j\big)=\int_0^1 \frac{|C_{f_i}(t)-C_{f_j}(t)||f_i(t)-f_j(t)|}{\lambda + \sqrt{H_{f_i}(t)H_{f_j}(t)}}dt.
\end{eqnarray}
 In eq.~(\ref{p2pmetric}), $\lambda$ is a positive constant to avoid singularity.

\subsection{Elastic morphing and SRVF}
\label{emorph}
The metric in (\ref{p2pmetric}) contains the absolute difference term $|f_i(t)-f_j(t)|$, which is the Euclidean distance between the two paths. From the point of view of geometry, a path can be thought of as an open curve which starts from a designated root node, and ends at a given dendritic terminal. However, the submanifold $\subset \mathcal{L}^2([0,1],\mathcal{R}^3)$ consisting of all the open and closed curves is not Euclidean. To measure the geodesic distance between $f_i$ and $f_j$, we need a suitable shape representation and a Riemannian metric. 

It is shown in~\cite{srivastava2011shape} that after the transformation of the space of curves by square root velocity (SRV) function, the space becomes Euclidean with the Euclidean distance acting as an elastic metric. The SRV of an arbitrary path $f_i$ can be given by $q_i(t) = \dot{f}_i(t)/\sqrt{||\dot{f}_i(t)||};~t\in[0,1]$. 
The \emph{elasticity} comes from the fact that a curve can continuously deform from one shape to another like a rubber band. To perform such continuously elastic morphing, one needs to take care of scaling variability, rotation, translation, and reparametrization of curves. 

\emph{\textbf{Translation}:} The SRV transformation inherently takes care of the translation factor. 

\emph{\textbf{Scaling}:} To tackle the scaling variability, authors in~\cite{srivastava2011shape} restrict the lengths of all the curves to unity, which transforms the flat Euclidean space to a sphere. Therefore, in order to retrieve the intermediate deformations from one curve to another, one needs to find and traverse the geodesic path on the hypersphere. This poses a problem in the path matching between neurons. The restriction of unit length significantly alters the morphology of the paths. This is because the paths from the root to the dendritic terminals of a neuron differ in lengths, creating the distinctive morphology of the neuron. In our work, we do not impose the restriction of unit length of a path.

\emph{\textbf{Rotation and Reparametrization}:} The rotation group, $SO(3)$ and the reparametrization group, $\Gamma$ are compact Lie groups. Let $M$ be the Euclidean manifold of all the open curves after SRV transformation. The individual quotient spaces, $M/SO(3)$ and $M/\Gamma$ are also submanifolds inheriting the Riemannian metric of $M$, which ensures that the quotient space, $M/\big(\Gamma\times SO(3)\big)$ is also a submanifold. Therefore, any arbitrary path $f_i\in M$ is, at first, subjected to rotation and reparametrization, if required, followed by the SRV transformation $q_i$ to find an element in the quotient space.  
The registration of $f_i$ with respect to $f_j$ via rotation is performed by Kabsch algorithm~\cite{kabsch1978discussion}, which registers two sets of coordinate vectors. 

To introduce reparametrization, first note that the numerical implementation of (\ref{p2pmetric}) requires an equal number of vertices in $f_i$ and $f_j$. However, in practice, the number of vertices differs significantly from path to path. In addition, the locations of the vertices are fairly nonuniform to account for the path fragmentation, \emph{wiggliness} of path segments~\cite{batabyal2018neurobfd}, which are essential structural characteristics of a neuron. It is evident that for two curves of arbitrary lengths, more samples (vertices) approximate (\ref{p2pmetric}) as an integral. The error in distance, $|f_i(t)-f_j(t)|$ between two curves decreases with the increase in the number of samples. Notice that the resampling of a path is a class of reparametrization function.

In contrast to the resampling routine in Path2Path, we keep the positions of the actual vertices on a path $f_i$ fixed, and add other samples in between in an iteratively sequential fashion. Between two consecutive samples on a path, we insert the midpoint of the samples as a new point. This procedure retains the neuronal characteristics of each path. The concurrence, $C_{f_i}, C_{f_j}$ and the hierarchy, $H_{f_i}, H_{f_j}$ values are interpolated as $\tilde{C}_{f_i}, \tilde{C}_{f_j}$ and $\tilde{H}_{f_i}, \tilde{H}_{f_j}$ respectively. 
 
Let the number of vertices of $f_i$ and $f_j$ are $N_i$ and $N_j$. The number of samples in the resampling procedure is fixed as $\rho$. After reparamterization, the paths become $\tilde{f}_i$ and $\tilde{f}_j$, which are subjected to SRV function producing $q_i$ and $q_j$ respectively. The $q_i$ is then rotated as $\tilde{q}_i$ with respect to $q_j$ for registration. The distance $D\big(f_i,f_j\big)$ between the two paths is computed by inserting $\tilde{q}_i$, $q_j$, $\tilde{C}_{f_i}, \tilde{C}_{f_j}$, $\tilde{H}_{f_i}$ and$\tilde{H}_{f_j}$ in (\ref{p2pmetric}).   
\vspace{-0.4cm}
\subsection{Path-to-Path matching}
\label{p2pm}
\vspace{-0.2cm}
In Path2Path~\cite{basu2011path2path}, the matching algorithm is greedy, which has a serious drawback of singularity in which all paths in one neuron can be matched to only one path in the other neuron. We tackled the problem by defining a one-to-one job assignment problem. 
Let us consider two different neurons, $\mathcal{G}_1$ and $\mathcal{G}_2$ having the sets of paths as $P_1 = \{f_1^1, f_2^1,\dots f_{|P_1|}^1\}$ and $P_2 = \{f_1^2, f_2^2,\dots f_{|P_2|}^2\}$ with $|P_1|\le |P_2|$ respectively. Here $|P|$ indicates the cardinality of the set $P$. The distance measure, $D(f_i^1, f_j^2)$ as defined in eq.~(\ref{p2pmetric}) can be regarded as a cost between two paths $f_i^1$ and $f_j^2$. Let $C\in\mathcal{R}^{|P_1|\times|P_2|}$ be the matrix with $C(i,j) = D(f_i^1, f_j^2)$. 
The path-to-path matching problem between two neurons can be regarded as a variant of a job assignment problem. 
Here, $|P_1|$ is the number of workers and $|P_2|$ is the number of jobs that are to be assigned to the workers. As in most of the cases, $|P_1|\neq|P_2|$, the assignment problem is unbalanced. We append $(|P_2|-|P_1|)$ zero rows to the bottom of $C$ as dummy workers. The optimal one-one job assignment is then performed using the Hungarian algorithm~\cite{munkres1957algorithms}.      
\vspace{-0.4cm}
\section{Datasets and Results}
We apply Elastic Path2Path (ElasticP2P for short) on a dataset containing (.swc format) files of digitally-traced neurons taken from Neuromorpho.org. The dataset consists of five major cell types - pyramidal, granule, motor, purkinje, and ganglion. To restrict the model organism, we consider the murine neurons only. There are total 4434 neurons used in our study, out of which 1729 are pyramidal, 1195 are granule, 116 are motor, 57 are purkinje, and 1377 are ganglion. 
\begin{figure}[ht]
\vspace{-0.4cm}
\centering
	\subfigure[]{\includegraphics[width=4.2cm,height=2.7cm]{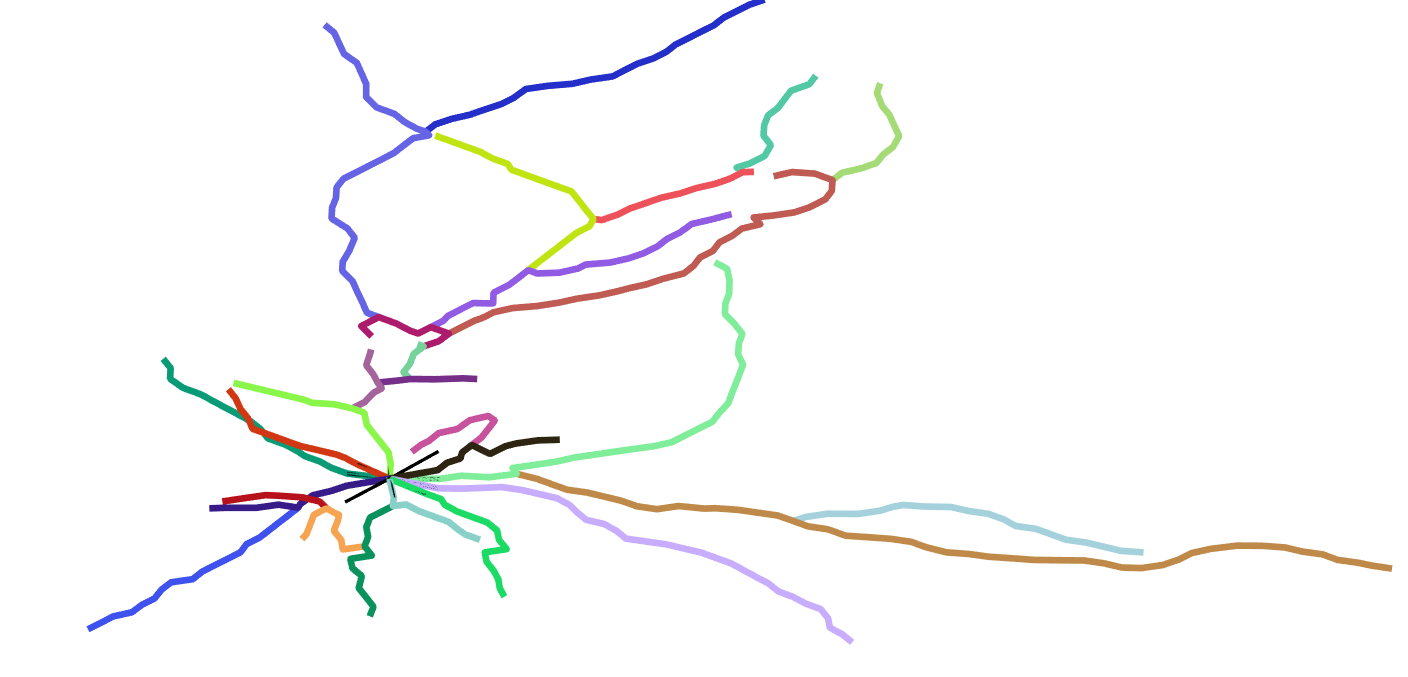}}
	\subfigure[]{\includegraphics[width=4.2cm,height=2.7cm]{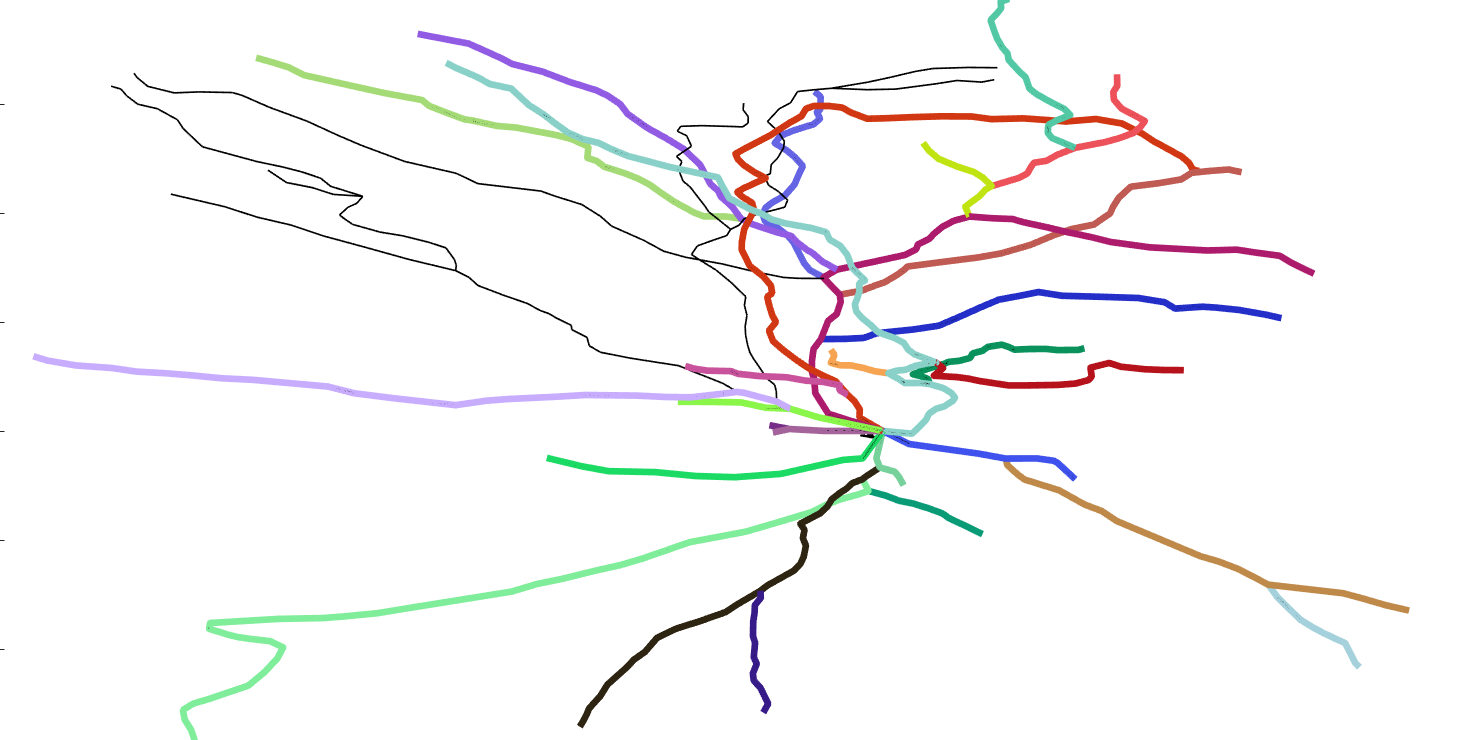}}
	\subfigure[]{\includegraphics[width=4.2cm,height=2.7cm]{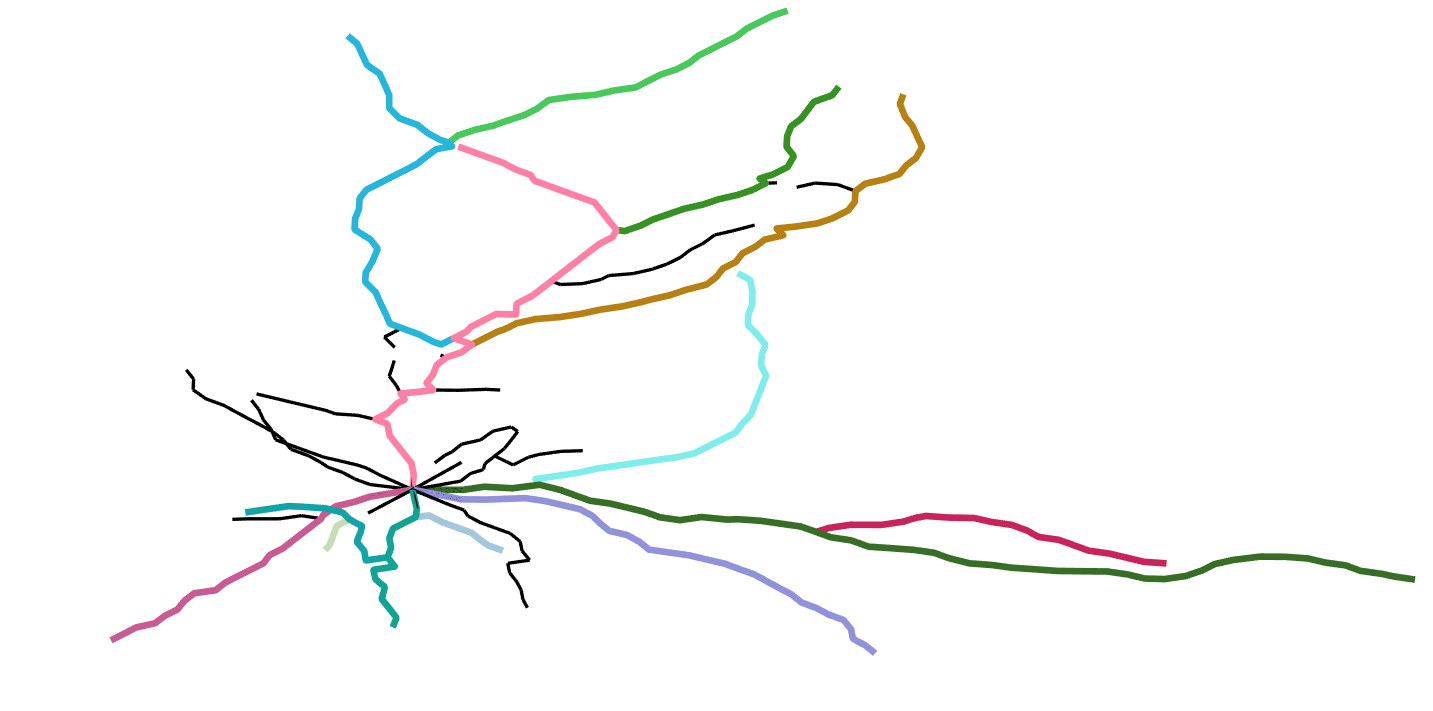}}
	\subfigure[]{\includegraphics[width=4.2cm,height=2.7cm]{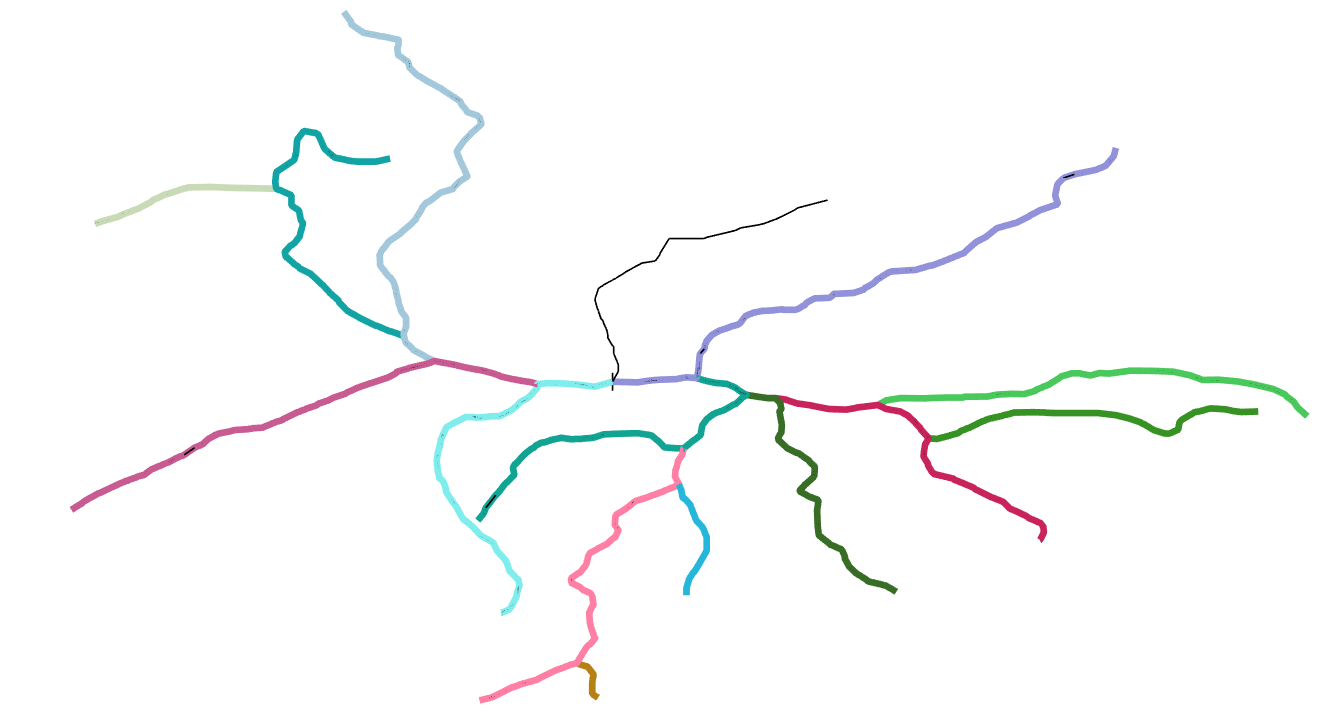}}
	\subfigure[]{\includegraphics[width=4.2cm,height=2.7cm]{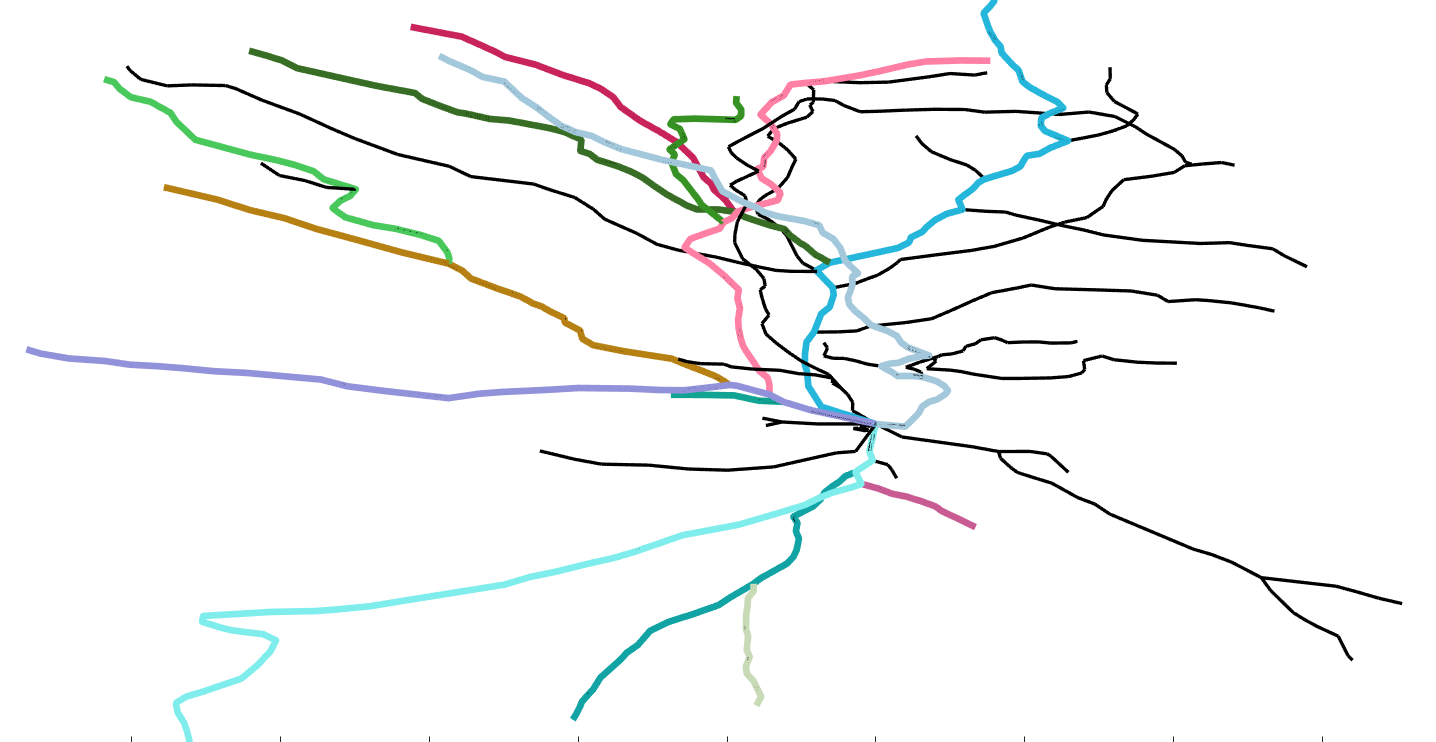}}
	\subfigure[]{\includegraphics[width=4.2cm,height=2.7cm]{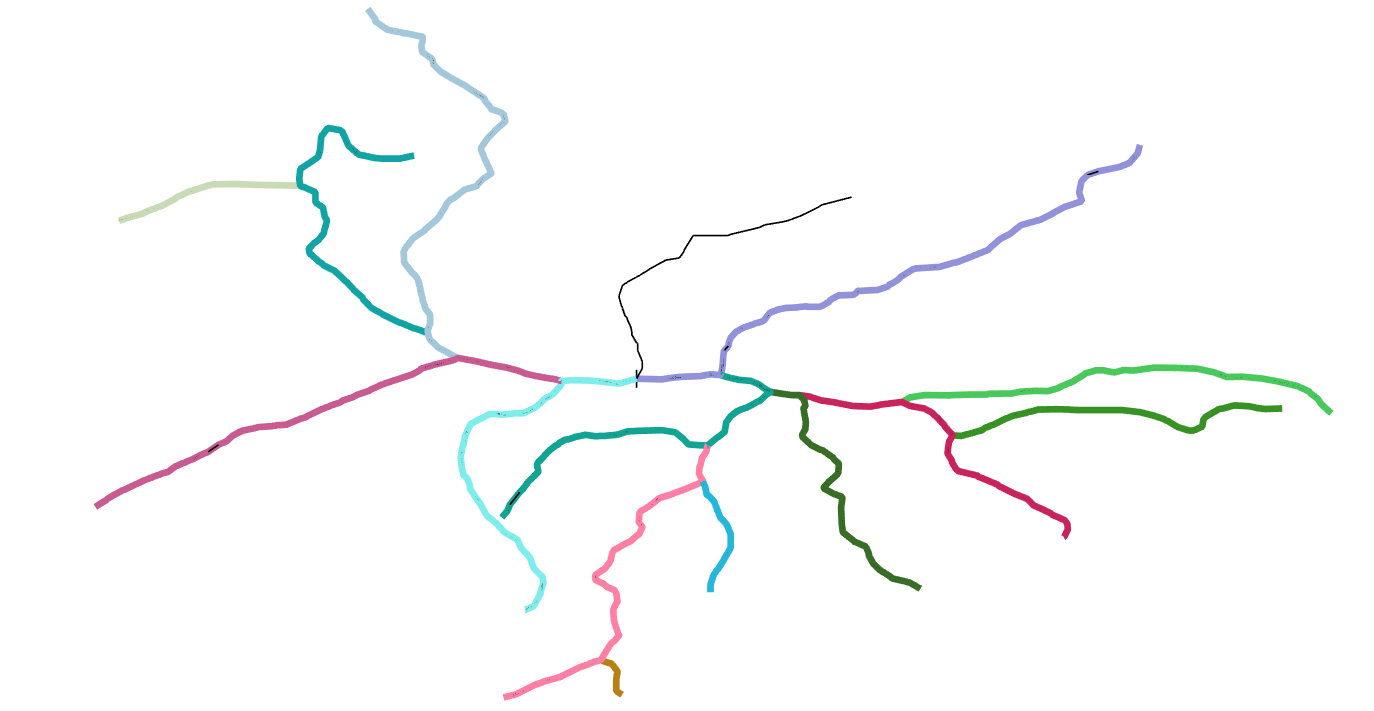}}
	\caption{\small The path correspondences are shown in the corresponding colors in case of (a)-(b) pyramidal-motor, (c)-(d) pyramidal-ganglion, and (e)-(f) motor-ganglion cell types. The paths which are matched are depicted in thick colored lines. The paths which are left out (larger neurons) are shown in thin black lines. }
\label{fig2}
\vspace{-0.3cm}
\end{figure}
An example of the path correspondences in ElasticP2P in case of three major cell types, which are pyramidal, motor, and ganglion are exhibited in Fig.~\ref{fig2}. The pyramidal, motor, and ganglion neuron samples contain $28$, $35$, and $14$ rooted paths respectively. The costs of morphing from one neuron to another neuron are 478.67 (pyramidal-motor pair), 514.09 (pyramidal-ganglion pair), and 353.14 (motor-ganglion pair). 
We also check the consistency of ElasticP2P by computing distances between same samples, which turn out to be null.

\begin{figure}[ht]
\vspace{-0.5cm}
\centering
	\includegraphics[width=8.2cm,height=3.1cm]{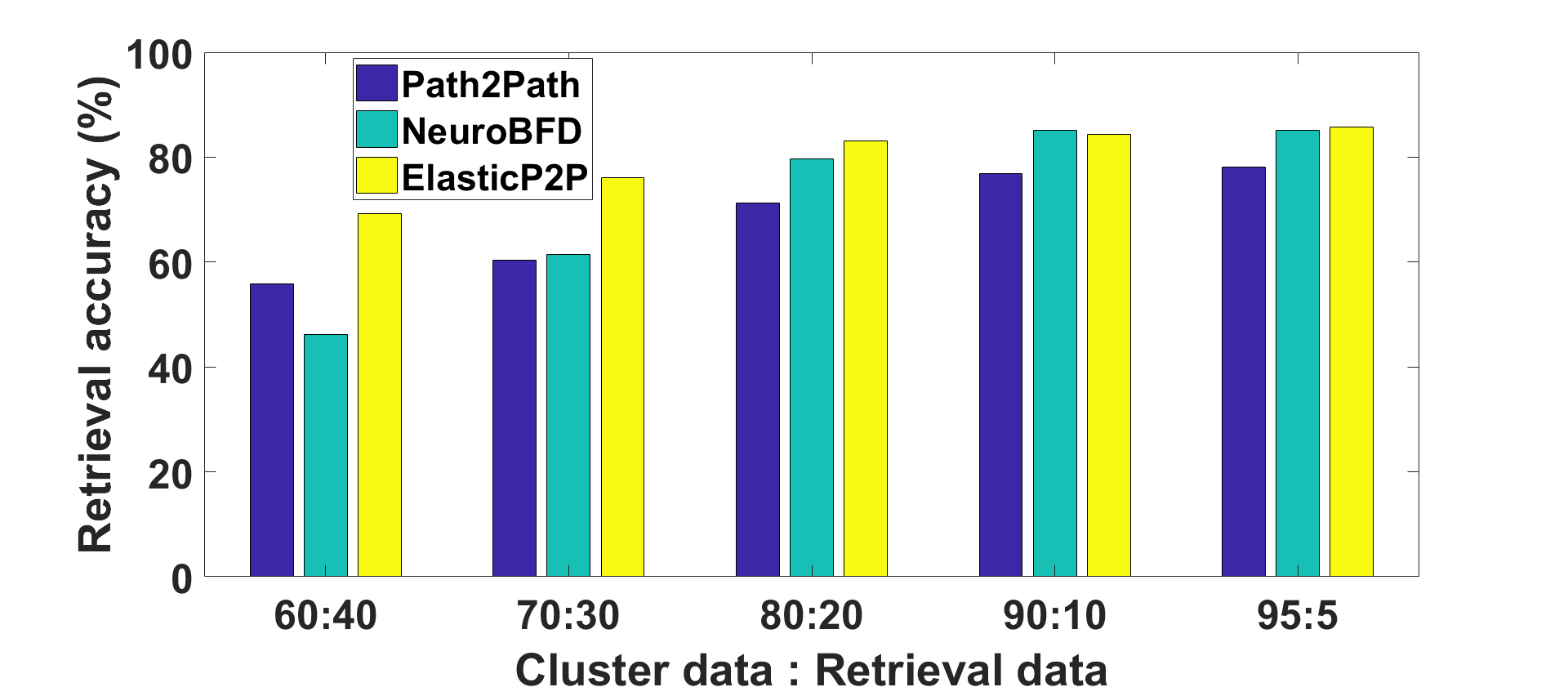}
	\caption{\small The retrieval performance (in \%) of ElasticP2P against Path2Path~\cite{basu2011path2path} and NeuroBFD~\cite{batabyal2018neurobfd}.}
\label{fig3}
\vspace{-0.5cm}
\end{figure}

Due to the variation in the cardinalities of the sets for the five cell types, we resort to unsupervised classification. The dataset is partitioned into a set for clustering and a test set for determining the retrieval accuracy. We perform different levels of partitioning of  the dataset to test the resilience of our approach over NeuroBFD and Path2Path. At each level, we partitioned the dataset five times randomly maintaining the same ratio between the size of the training and test dataset. 
The retrieval is carried out using majority vote rule. At each partition, for a candidate in the retrieval set, we compute the nearest $11$ samples from the cluster set. The class which appears largest number of times out of $11$ labels is assigned to the candidate neuron. 
We average the retrieval scores and show them in Fig.~\ref{fig3}. In all the instances of this experiment, $\rho$ is kept $100$. 
The results suggest that EasticP2P exhibits consistent performance over a wide range of data partitioning compared to Path2Path and NeuroBFD. The classification method in NeuroBFD is supervised, which used a set of linear SVMs. With the gradual reduction in the amount of training data, the performance of NeuroBFD collapses as evident from Fig.~\ref{fig3}. At the data ratio of $9:1$, NeuroBFD performs marginally better ($85.1\%$) than ElasticP2P ($84.3\%$). The consistent improvement over Path2Path can be attributed to the insertion of the one-to-one path assignment routine and the rectified resampling routine in ElasticP2P.   

\begin{figure}[ht]
\vspace{-0.3cm}
\centering
	\includegraphics[width=8.2cm,height=3.1cm]{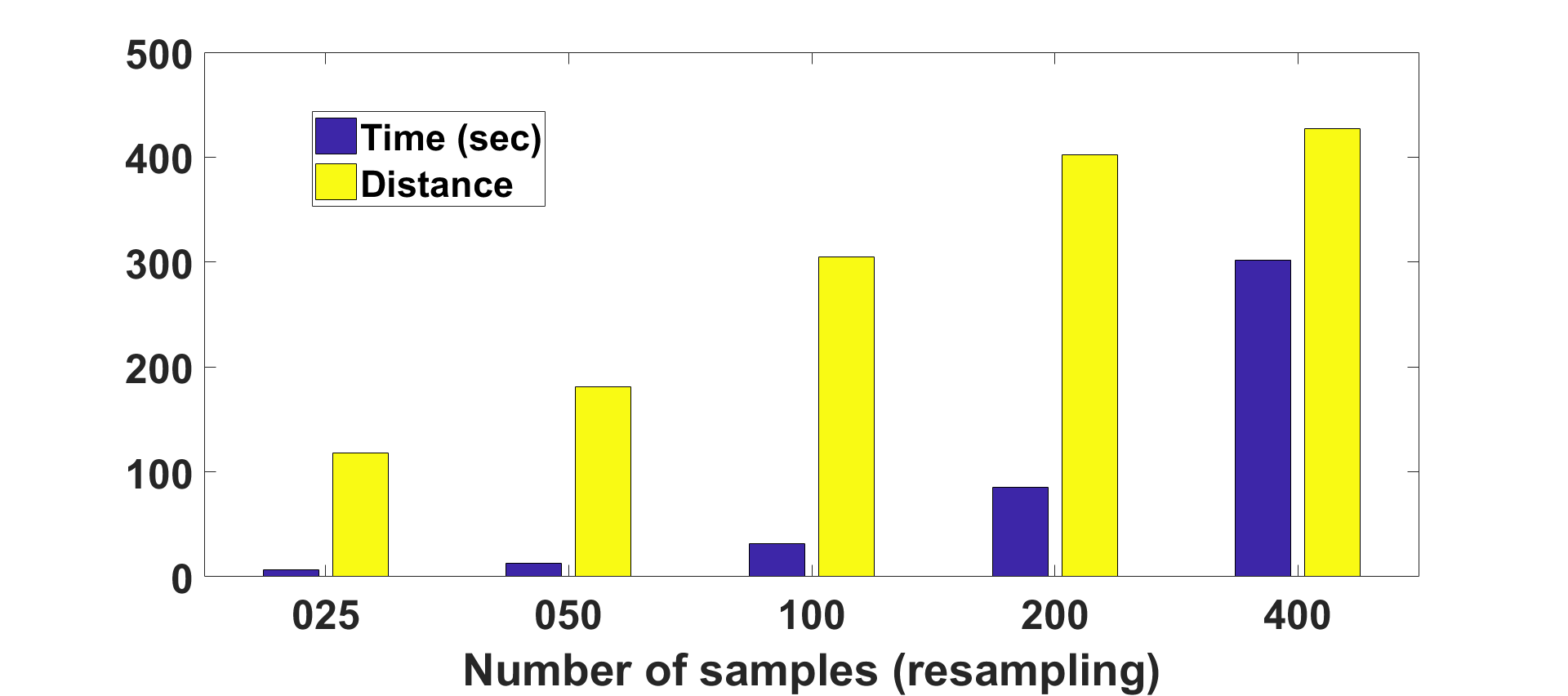}
	\caption{\small The plot shows the distance between a pyramidal and a ganglion neuron and the computational time (in seconds) when 25, 50, 100, 200 and 400 samples are taken per path.}
\label{fig4}
\vspace{-0.2cm}
\end{figure}

The only hyperparameter of ElasticP2P is $\rho$, the number of samples after resampling. We demonstrate the effectiveness in terms of retrieval accuracy and computational demand in terms of execution time of our algorithm with different choices of $\rho$ in Fig.~\ref{fig4}.
In Fig.~\ref{fig4}, a pyramidal cell and a ganglion cell containing 28 rooted paths with 463 3D locations and 40 paths with 2588 3D locations respectively are selected. The distance and computation time required as a function of the number of samples after resampling are shown. 
\vspace{-0.4cm}
\section{conclusion}
\vspace{-0.4cm}
ElasticP2P demonstrates superior performance in terms of the retrieval accuracy, reducing intra-cell-type distances and enhancing inter-cell-type distances. In addition, it provides \emph{visual evidence} of how a pair of neurons is compared in terms of path correspondences, which makes ElasticP2P a methodology usable in the biology laboratory. The graph theoretic representation of a neuron is intertwined with a differential geometric framework to allow the continuous morphing between a pair of neurons. In the future, we intend to use this tool to study the behavior of microglia by exploiting their dynamic morphology. We are also interested in the quantitative assessment of the slow but continuous structural degeneration of neurons, such as observed in Alzheimer's disease.    

\bibliographystyle{IEEEbib}
\bibliography{elasticP2P}
\end{document}